\documentstyle[12pt]{article} 

\def\beq{\begin{equation}}  
\def\eeq{\end{equation}}  
\def\beqa{\begin{eqnarray}}  
\def\eeqa{\end{eqnarray}}   
  
\begin{document}  
   
\begin{titlepage}  
  
\begin{flushright}  
{\sc UMHEP-452}\\ [.2in]  
{\sc May 7, 1998}\\ [.5in]  
\end{flushright}  
  
\begin{center}  
  
{\LARGE  
Form factor relations for pseudoscalar to vector meson transitions.}\\
[.5in]
{\large Jo\~{a}o M. Soares}\\ [.1in]  
{\small
Department of Physics and Astronomy\\  
University of Massachusetts\\  
Amherst, MA 01003-4525}\\ [.5in]

{\normalsize\bf   
Abstract}\\ [.2in]  

\end{center}  

{\small
In semi-leptonic and other weak decays of mesons, the hadronic matrix elements 
of the operators in the weak Hamiltonian are parametrized by standard sets of 
independent, Lorentz invariant, form factors. For the case of pseudoscalar to 
vector meson transitions, it has been shown that a Quark Model description 
leads to relations between some of the form factors. Here, I give an alternate,
and more general, proof of those relations and thus confirm that, in the Quark
Model, not all the form factors for pseudoscalar to vector meson transitions 
are independent. As an application of this result, a Quark Model measurement 
of the CKM parameter $|V_{ub}|$ can be obtained, where the dependence on all 
hadronic form factors is eliminated.
}   
  
\end{titlepage}

\section{Introduction}  
 
The study of the weak decays of quarks, and the measurement of the 
corresponding CKM matrix elements, are consistently hampered by the presence 
of the long distance QCD effects that are responsible for the binding of the 
quarks into hadrons. These effects are hard to evaluate in a model independent
way, and so tend to bring large uncertainties to the theoretical predictions 
for the weak decay amplitudes. They appear in the calculation of the matrix 
elements of the weak Hamiltonian operators, between the initial and final 
hadronic states. For practical purposes, they are included in standard sets of
independent, Lorentz invariant, form factors, that parametrize those hadronic 
matrix elements in a convenient way, but are otherwise poorly known. Relations
between different form factors, that will hold under certain conditions or 
approximations, can then be very useful: they will reduce the number of 
uncertain quantities, and improve the accuracy of the theoretical predictions.
Moreover, they may help us understand better the general features of the 
underlying long distance QCD effects.

Here, I am interested in the weak hadronic matrix elements for a pseudoscalar
$B$ to a vector meson $V$ transition (for definiteness, I use a $B$-meson in 
my notation, but the results are quite general; they apply equally well to 
$D$-mesons, or even to lighter pseudoscalars, if they were to decay weakly to
vector mesons). These matrix elements are parametrized by the form factors 
$V$, $A_{0,1,2}$ and $F_{1,2,3}$, that are defined as follows:
\beqa 
\lefteqn{\langle V(\vec{p^\prime},\vec{\varepsilon})|  
\overline{q} \gamma^\mu b |B(\vec{p}) \rangle  = 
\frac{-1}{m_B + m_V} 2 i \epsilon^{\mu\alpha\beta\gamma} 
\varepsilon_\alpha^\ast p_\beta^\prime p_\gamma V(k^2) \ ,}  
\label{0}\\
\nonumber\\
\lefteqn{\langle V(\vec{p^\prime},\vec{\varepsilon})|  
\overline{q} \gamma^\mu \gamma_5 b |B(\vec{p}) \rangle} \nonumber\\
& &  = (m_B + m_V) \left( \varepsilon^{\ast\mu} -   
\frac{\varepsilon^\ast.k}{k^2} k^\mu \right) A_1(k^2) \nonumber\\   
& & - \frac{\varepsilon^\ast.k}{m_B + m_V} 
\left( (p+p^\prime)^\mu - \frac{m_B^2 - m_V^2}{k^2} k^\mu \right) A_2(k^2) 
\nonumber\\ 
& & + 2 m_V \frac{\varepsilon^\ast.k}{k^2} k^\mu A_0(k^2) \ ,  
\label{1}\\ 
\nonumber\\ 
\lefteqn{\langle V(\vec{p^\prime},\vec{\varepsilon})|   
\overline{q} i \sigma^{\mu\nu} k_\nu b |B(\vec{p}) \rangle 
= i \epsilon^{\mu\alpha\beta\gamma} 
\varepsilon_\alpha^\ast p_\beta^\prime p_\gamma F_1(k^2) \ ,} 
\label{2}\\  
\nonumber\\
\lefteqn{\langle V(\vec{p^\prime},\vec{\varepsilon})|   
\overline{q} i \sigma^{\mu\nu} \gamma_5 k_\nu b |B(\vec{p})\rangle 
\frac{}{}} 
\nonumber\\ 
& & = \left( (m_B^2 - m^2_V) \varepsilon^{\ast\mu}    
- \varepsilon^\ast.k (p+p^\prime)^\mu \right) F_2(k^2) 
\nonumber\\   
& & + \varepsilon^\ast.k \left( k^\mu - \frac{k^2}{m_B^2 - m^2_V}    
(p+p^\prime)^\mu \right) F_3(k^2) \ ,  
\label{3}  
\eeqa 
with $2 m_V A_0(0) = (m_B + m_V) A_1(0) - (m_B - m_V) A_2(0)$, $F_1(0) =  
2 F_2(0)$ and $k \equiv p - p^\prime$.

In Ref.\ \cite{JMS}, a Quark Model description of the $B \to V$ hadronic 
transition led to the following relations between the form factors $F_{1,2,3}$ 
and $A_{0,1,2}$:
\beqa 
F_1(k^2) &=& 2 A_0(k^2) \ ,  
\label{4}\\ 
\nonumber\\ 
m_V (m_B - m_V) F_2(k^2) &=& (m_B E^\prime - m_V^2) A_1(k^2) 
\nonumber\\ 
& & - \frac{2 m_B^2 |\vec{p^\prime}|^2}{(m_B + m_V)^2} A_2(k^2) \ , 
\nonumber\\ 
\label{5}\\ 
(m_B E^\prime + m_V^2) F_2(k^2)
&-& \frac{2 m_B^2 |\vec{p^\prime}|^2}{m_B^2 - m_V^2} F_3(k^2) 
\nonumber \\ 
&=& m_V (m_B + m_V) A_1(k^2) \ .
\label{6}  
\eeqa 
Note that these relations are valid in any reference frame; the energy 
$E^\prime$ and momentum $|\vec{p^\prime}|$ of the vector meson $V$, in the 
rest frame of the $B$-meson, are used as an abbreviation for the more 
cumbersome invariant functions of $k^2$:
\beqa  
E^\prime &=& \frac{m_B^2 + m_V^2 - k^2}{2 m_B} \ , \\  
\label{7}  
|\vec{p^\prime}| &=& \frac{\left[ (m_B^2 + m_V^2 - k^2)^2  
- 4 m_B^2 m_V^2 \right]^{1/2}}{2 m_B}  \ .  
\label{8}  
\eeqa  
The fact that these relations were derived, independently of the details of 
the momentum wavefunctions for the two mesons, suggests that they are a very
general Quark Model result. However, the proof in Ref.\ \cite{JMS} was not 
entirely satisfactory, in that the origin of these form factor relations was 
not clear, and the outcome seemed rather accidental. In what follows, I will 
give an alternate and more general proof of the same relations. It will 
clarify their origin, and the Quark Model assumptions in which they rely. The 
end result will confirm that, in the Quark Model, the form factors $F_{1,2,3}$ 
for pseudoscalar to vector meson transitions are not independent form factors.

\section{Form factor relations}  
 
The first step in deriving the form factor relations is to solve 
eqs.~\ref{0}--\ref{3}, in order to write each form factor as a combination of 
hadronic matrix elements. This is most easily done in the $B$ rest frame, with
the $z$-axis chosen along the momentum $\vec{p^\prime}$ of the vector meson 
$V$. In this frame,
\beq 
p = m_B (1, 0, 0, 0)  \;\;\;\;\;\;\; 
p^\prime = (E^\prime, 0, 0, |\vec{p^\prime}|)  \ ,
\label{9}
\eeq 
and the polarization vectors for the helicity states $\lambda = 0$, $\pm 1$
of the vector meson are 
\beq
\varepsilon_0 = \frac{1}{m_V} (|\vec{p^\prime}|, 0, 0, E^\prime) \;\;\;\;\;\; 
\varepsilon_\pm = \mp \frac{1}{\sqrt{2}} (0, 1, \pm i, 0) \ ;
\label{10} 
\eeq
$k \equiv p - p^\prime = (m_B - E^\prime, 0, 0, -|\vec{p^\prime}|)$, and it is 
convenient to define a 4-vector $r = N (|\vec{p^\prime}|, 0, 0, -m_B + 
E^\prime)$, with arbitrary normalization, such that $r.k = 0$ and $\vec{r}$ is
parallel to $\vec{k}$, in the particular frame that we have chosen. The 
expressions that are obtained for each of the form factors can then be 
combined as follows:
\beqa
\lefteqn{F_1(k^2) = 2 A_0(k^2) \;\; 
\frac{k_\mu \langle V(\lambda = \pm 1) | \overline{q} \frac{i}{\sqrt{2}}  
( \sigma^{1 \mu} \pm i \sigma^{2 \mu} ) b | B \rangle}   
{k_\mu \langle V(\lambda = 0) | \overline{q} \gamma^\mu \gamma_5 b 
| B \rangle}  \ ,}
\label{12}\\
\nonumber\\
\lefteqn{m_V (m_B - m_V) F_2(k^2)} \nonumber\\
& & = \left[ (m_B E^\prime - m_V^2) A_1(k^2)
- \frac{2 m_B^2 |\vec{p^\prime}|^2}{(m_B + m_V)^2} A_2(k^2) \right] 
\nonumber \\
& & \times \;\; \frac{r_\mu \langle V(\lambda = \pm 1) | \overline{q} 
\frac{i}{\sqrt{2}} ( \sigma^{1 \mu} \pm i \sigma^{2 \mu} ) b | B \rangle}   
{r_\mu \langle V(\lambda = 0) | \overline{q} \gamma^\mu \gamma_5 b | B
\rangle} \ ,
\label{13}\\
\nonumber\\
\lefteqn{(m_B E^\prime + m_V^2) F_2(k^2)
- \frac{2 m_B^2 |\vec{p^\prime}|^2}{m_B^2 - m_V^2} F_3(k^2)}
\nonumber \\ 
& & = m_V (m_B + m_V) A_1(k^2) \; \nonumber\\
& & \times \;\; \frac{\varepsilon^\pm_\mu \langle V(\lambda = 0) 
| \overline{q} \frac{i}{\sqrt{2}} ( \sigma^{1 \mu} \mp i \sigma^{2 \mu} ) 
b | B \rangle}   
{\varepsilon^\pm_\mu \langle V(\lambda = \pm 1) | \overline{q} 
\gamma^\mu \gamma_5 b | B \rangle}  \ . 
\label{14}
\eeqa
The remainder of the proof consists in showing that the ratios of hadronic
matrix elements, that appear on the right-hand-side (RHS) of 
eqs.~\ref{12}--\ref{14}, are precisely equal to $1$, in the Quark Model. 

In order to do so, we must relate the $\lambda = 0$ and $\lambda = \pm 1$ 
helicity states of the vector meson. If one is to rely on the Quark Model 
angular momentum wavefunction for the vector meson, this can be done by 
flipping the spin of the constituent quark $q$: 
\beqa 
\langle V(\lambda = \pm 1) | \overline{q} \Gamma b | B \rangle &=& 
\sqrt{2} \langle V(\lambda = 0) | \; S_\pm^\dag \; \overline{q} \Gamma b 
| B \rangle \ , 
\label{15} 
\eeqa 
where $S_\pm$ are the raising and lowering operators for the spin of the quark
$q$, along the direction of the vector meson momentum; $\Gamma$ is an 
arbitrary combination of Dirac $\gamma$-matrices. The spin operator, for the 
$q$-quark field, is \cite{I&Z} 
\beqa 
S_\sigma &=& \frac{1}{2 m} \epsilon_{\sigma\mu\nu\rho} J^{\mu\nu} 
P^\rho  \ ,
\label{16}
\eeqa
where
\beqa   
J^{\mu\nu} &=& \int d^3x : q^\dag(x) \left[ i x^\mu D^\nu  - i x^\nu D^\mu
+ \frac{1}{2} \sigma^{\mu\nu} \right] q(x) :  
\label{17}  
\eeqa
and
\beqa   
P^\mu &=& \int d^3x : q^\dag(x) i D^\mu q(x) :  
\label{18}
\eeqa
are the conserved angular and linear momentum operators; they satisfy the 
commutation relations
\beqa 
[J^{\mu\nu}, q(x)] &=& - ( i x^\mu D^\nu - i x^\nu D^\mu  
+ \frac{1}{2} \sigma^{\mu\nu} ) q(x)  
\label{19}  
\eeqa  
and   
\beqa   
[P^\mu, q(x)] &=& - i D^\mu q(x)  \ .
\label{20}  
\eeqa
The covariant derivative $D^\mu$ accounts for the strong interactions of the 
quark field; it also appears in the equation of motion: $(i \not\!\!D - m) 
q(x) = 0$. The spin projection operator along a direction $n$, with $P.n=0$ 
and $n^2=-1$, is $S.n$, and it is then easy to arrive at the central relation 
in our proof,
\beqa
\langle V | (S.n)^\dag \overline{q} \Gamma b | B \rangle &=& 
\frac{1}{2} \langle V | \overline{q} \gamma_5 \gamma_\sigma \Gamma b  
| B \rangle n^{\sigma\ast} \ .
\label{21}
\eeqa
In the derivation, we assumed, as in the Quark Model, that the $B$-meson state
does not contain the quark $q$; then, $J^{\mu\nu}| B \rangle = P^\mu | B 
\rangle = 0$.
 
In order to apply this result to the matrix element on the RHS of 
eq.~\ref{15}, one must proceed with care. The spin projection operator along 
the z-axis, $S_z \equiv S.n^{(3)}$, and  the spin raising and lowering 
operators, $S_\pm \equiv S.(n^{(1)} \pm i n^{(2)})$, take a simple form in the
$q$-quark rest frame, where
\beq 
n^{(1)} = (0,1,0,0) \ , \;\;\; 
n^{(2)} = (0,0,1,0) \;\;\; {\rm and} \;\;\;  
n^{(3)} = (0,0,0,1) \ ; 
\label{22} 
\eeq 
then, $S_z = S_3$ and $S_\pm = S_1 \pm i S_2$. In the frame that we have 
chosen, with the B meson at rest and the vector meson momentum along the 
z-axis, the direction vectors $n^{(k)}$ depend on the $q$-quark momentum in 
that frame, and so the spin projection operators do not have, in general, such
a simple form. If, however, we can neglect the transverse momentum of the 
$q$-quark inside the vector meson $V$, then $n^{(1)}$ and $n^{(2)}$ are
as before. Under that assumption, eqs.~\ref{15} and \ref{21} give
\beqa 
\langle V(\lambda = \pm 1) | \overline{q} \Gamma b | B \rangle &=& 
\frac{1}{\sqrt{2}} \langle V(\lambda = 0) | \overline{q} \gamma_5  
( \gamma_1 \mp i \gamma_2 ) \Gamma b | B \rangle \ . 
\label{23} 
\eeqa 
This relation is now applied to the hadronic matrix elements on the RHS of 
eqs.~\ref{12}--\ref{14}, where $\Gamma = i/\sqrt{2} (\sigma^{1 \mu} \pm i 
\sigma^{2 \mu}) k_\mu$, $i/\sqrt{2} (\sigma^{1 \mu} \pm i \sigma^{2 \mu}) 
r_\mu$ or $\varepsilon^\pm_\mu \gamma^\mu \gamma_5$. It is straightforward 
to check that the ratio of hadronic matrix elements, in each one of the 
equations, is exactly $1$, and so we recover the form factor relations of 
eqs.~\ref{4}--\ref{6} and Ref.~\cite{JMS}.

\section{Measuring $|V_{ub}|$ in the Quark Model}

One possible application of the form factor relations derived in here is to 
the extraction of the CKM parameter $|V_{ub}|$, from the $B \to \rho l^-
\overline{\nu}_l$ decay rate. Since this exclusive decay rate depends on the 
form factors $V$ and $A_{0,1,2}$, that parametrize the matrix element $\langle
\rho | \overline{u} \gamma^\mu (1 - \gamma_5) b | B \rangle$, a clean
measurement of $|V_{ub}|$ is problematic \cite{IW}. However, we have seen 
that, in the Quark Model, these form factors are not independent from the form
factors $F_{1,2,3}$; they are the form factors that parametrize the matrix 
element $\langle \rho | \overline{d} i \sigma^{\mu\nu} k_\nu (1 + \gamma_5) b 
| B \rangle$, that appears, for example, in the amplitude for the radiative 
decay $B \to \rho \gamma$. A judicious comparison between the semi-leptonic 
and radiative decays can then yield $|V_{ub}|$, free from any form factor 
dependence \cite{BD}.

From the ratio of the exclusive $B \to \rho \gamma$ and the inclusive $B \to 
\gamma + X$ decay rates, 
\beqa
R_\rho &\equiv& \frac{\Gamma(B \to \rho \gamma)}{\Gamma(B \to \gamma + X)} 
\nonumber\\
&=& \left| \frac{V_{td}}{V_{ts}} \right|^2
\left( 1 - \frac{m_\rho}{m_B}\right)^3  
|\frac{1}{2} F_1(0)|^2 \ ,
\label{24}
\eeqa
we can obtain a measurement of the form factor $F_1(k^2)$ at $k^2 = p_\gamma^2
= 0$. On the other hand, the $B \to \rho l^- \overline{\nu}_l$ differential 
decay rate, at the $k^2 = (p_{l^-} + p_{\overline{\nu}_l})^2 = 0$ boundary of 
the Dalitz plot, is
\beqa
\lim_{k^2 \to 0} \frac{d\Gamma(B \to \rho l^- \overline{\nu}_l)}
{d(k^2/m_B^2)} &=&
\frac{G_F^2}{192 \pi^3}
\left( 1 - \frac{m_\rho}{m_B}\right)^2 m_B^5
|V_{ub}|^2 |A_0(0)|^2 \ .
\label{25}
\eeqa
Using the $F_1(k^2) = 2 A_0(k^2)$ relation of eq.~\ref{4}, the dependence on
the form factors can be eliminated between eqs.~\ref{24} and \ref{25}:
\beqa
\frac{1}{R_\rho} \; \lim_{k^2 \to 0}
\frac{d\Gamma(B \to \rho l^- \overline{\nu}_l)}{d(k^2/m_B^2)} &=&
\frac{G_F^2}{192 \pi^3}  m_B^5 
\left| \frac{V_{ts}}{V_{td}} \right|^2 |V_{ub}|^2 \ .
\label{26}
\eeqa
A measurement of the CKM parameters can then be obtained that is free of the 
hadronic form factor contributions. One must stress, however, that such a 
measurement is dependent on the Quark Model assumptions that led to the form 
factor relations of the previous section. Nevertheless, within that model, it 
is a very general result, as it does not depend on the particular choice for 
the momentum wavefunctions of the mesons involved.

For the inclusive radiative decay rate in eq.~\ref{24}, I have taken $\Gamma(B
\to \gamma + X) \simeq \Gamma(b \to s \gamma)$, $m_b \simeq m_B$ and $m_s 
\simeq 0$; for the present purpose, these are sufficiently good 
approximations. In the exclusive $B \to \rho \gamma$ amplitude, corrections 
proportional to $|V_{ub} V^\ast_{ud}|$, due to the difference between $u$ and 
$c$ quark loops in the weak vertex, have been ignored. These corrections 
have been estimated to be small, when compared to the dominant $t$ quark loop 
contribution \cite{JMSb}. Alternatively, one can consider the decay $B \to 
K^\ast \gamma$, where such corrections are irrelevant. The analogue of 
eq.~\ref{26}, in that case, will not have the extraneous factor 
$|V_{ts}/V_{td}|^2$, but it will be valid only up to $SU(3)_{flavor}$ 
symmetry-breaking corrections to the form factors. The semi-leptonic decay 
rate has been observed by CLEO, with $BR(B^0 \to \rho^- l^+ \nu_l) = (2.5 \pm 
0.4 ^{+0.5}_{-0.7} \pm 0.5) \times 10^{-4}$ \cite{CLEOa}; more data will be 
necessary, in order to determine the differential decay rate in eq.~\ref{25}. 
The inclusive radiative decay has also been measured, with $BR(B \to \gamma + 
X) = (2.32 \pm 0.57 \pm 0.35) \times 10^{-4}$ \cite{CLEOb}. As for the 
exclusive radiative decay, only the Cabibbo favored $B \to K^\ast \gamma$ has 
been seen, with $BR(B \to K^\ast \gamma) = (4.2 \pm 0.8 \pm 0.6) \times 
10^{-5}$ \cite{CLEOc,CLEOd}. The branching ratio for $B \to \rho \gamma$ is 
expected to be about $20$ times smaller, whereas the present limit is $BR(B 
\to (\rho,\omega) \gamma)/ BR(B \to K^\ast \gamma) < 0.19$  at $90\%$ C.L. 
\cite{CLEOd}.

\section{Conclusion}

The Quark Model relations between the form factors $F_{1,2,3}$ and 
$A_{0,1,2}$, that parametrize the weak hadronic transitions from pseudoscalar 
to vector me\-sons, were obtained in here from a different, and more general, 
argument than that in the original derivation of Ref.\ \cite{JMS}. This 
alternate proof confirms that, in the Quark Model, $F_{1,2,3}$ are no longer 
independent form factors. Moreover, it reveals how these form factor relations
originate from the spin wavefunctions for the mesons, but do not depend (with 
one caveat) on the particular choice for the internal momentum wavefunctions. 
In that respect, they are a very general result of the Quark Model. One should
also stress that these relations are valid throughout the entire $k^2$ range 
of the hadronic transition, and for any value of the pseudoscalar and vector 
meson masses. As for the caveat regarding the momentum wavefunctions, it 
arises from the need to assume that the momentum of the quark produced at the 
weak vertex is nearly parallel to the momentum of the vector meson into which 
it hadronizes. This is a good approximation in at least two situations that, 
together, nearly exhaust all scenarios that may occur in practice. One is the 
case of heavy-to-heavy transitions, where the momentum of the heavy quark is 
roughly that of the corresponding heavy meson. The other is the case of 
heavy-to-light transitions, at sufficiently large recoil: the light quark 
carries a fraction of the large vector meson momentum that dominates over its 
small transverse momentum. As an example where the latter situation occurs, 
the form factor relations were applied to the extraction of the CKM parameter 
$|V_{ub}|$, from a comparison of the exclusive $B \to \rho l^- 
\overline{\nu}_l$ and $B \to \rho \gamma$ decays, along similar lines to the 
method that was suggested in Ref.~\cite{BD}.

The question that one would now like to answer is whether the form factor 
relations obtained in here remain valid, to some degree, beyond the Quark 
Model. In the case of heavy-to-heavy transitions, it is easy to check that the
model independent form factor relations, that follow from Heavy Quark Symmetry 
(HQS) \cite{HQS}, do lead to the relations derived in here. This means that 
the Quark Model result is indeed a true QCD result, in that limit; however, it
also means that it adds little to the well known, and more general, form 
factor relations of HQS. On the other hand, for heavy-to-light transitions, 
the Quark Model result provides a new set of form factor relations \cite{h2l}.
A comparison with lattice QCD may shed some light into their validity beyond 
the Quark Model. Hopefully, future lattice calculations will address the 
question and test these form factor relations, in the more general setting of 
their approximation. Ultimately, when sufficient data is accumulated for 
heavy-to-light decays, a comparison with the experimental results will decide 
whether the Quark Model prediction for the form factors provides a reasonable 
picture of QCD.

\section*{Acknowledgments} This research was funded in part by a grant from
the National Science Foundation. I wish to thank Jean-Bernard Zuber, John 
Donoghue, Eugene Golowich and Lincoln Wolfenstein for helpful discussions.


\begin{thebibliography}{abcd} 

\bibitem{JMS} 
J. M. Soares, Phys. Rev. {\bf D 54}, 6837 (1996). 
 
\bibitem{I&Z} 
C. Itzykson and J.-B. Zuber, {\em Quantum Field Theory} (McGraw-Hill, 
Singapore, 1985). 

\bibitem{IW}
See, however,
N. Isgur and M. B. Wise, Phys. Rev. {\bf D 42}, 2388 (1990),
for the use of Heavy Quark Symmetry in extracting $|V_{ub}|$ from a 
comparison between the $B \to \pi l^- \overline{\nu}_l$ and $D \to \pi l^- 
\overline{\nu}_l$ decays.

\bibitem{BD}
An alternative method to extract $|V_{ub}|$, from a comparison between 
semi-leptonic and radiative $B$ decays, can be found in
G. Burdman and J. F. Donoghue, Phys. Lett. {\bf B 270}, 55 (1991) .

\bibitem{JMSb}
J. M. Soares, Phys. Rev. {\bf D 49}, 283 (1994). 
 
\bibitem{CLEOa}
CLEO Collaboration, Phys. Rev. Lett. {\bf 77}, 5000 (1996).

\bibitem{CLEOb}
CLEO Collaboration, Phys. Rev. Lett. {\bf 74}, 2885 (1995).

\bibitem{CLEOc}
CLEO Collaboration, Phys. Rev. Lett. {\bf 71}, 674 (1993).

\bibitem{CLEOd} 
CLEO Collaboration, CLEO CONF 96-05.

\bibitem{HQS}
For a review of the Heavy Quark Symmetry results for the form factors, see,
for example, M. Neubert {\em et al.}, in {\em Heavy Flavors}, ed. by A. J.
Buras and L. Lindner (World Scientific, Singapore, 1992). 

\bibitem{h2l} 
For heavy-to-light transitions, model independent relations can be obtained 
in the static heavy quark limit \cite{IW}. The Quark Model results in here
cannot be derived from them, and they are indeed a new set of form factor 
relations.

\end{thebibliography}
\end{document}